\documentclass[preprint,floats,aps,superscriptaddress,showpacs]{revtex4}

\usepackage{amsmath}
\usepackage{epsfig}

\newcommand{\vp}{{\bf  v}}

\newcommand{\bx}{{\bf  x}}
\newcommand{\by}{{\bf  y}}
\newcommand{\br}{{\bf  r}}
\newcommand{\bg}{{\bf  g}}

\newcommand{\BE}{\begin{equation}}
\newcommand{\EE}{\end{equation}}
\newcommand{\beqn}{\begin{eqnarray}}
\newcommand{\eeqn}{\end{eqnarray}}

\newcommand{\etab}{\mbox{\boldmath $\eta$}}

\begin{document}

\title{Clustering transition in a system of  
particles self-consistently driven by a shear flow }

\author{Crist\'obal L\'opez}
\address{Instituto
Mediterr\'aneo de Estudios Avanzados IMEDEA (CSIC-UIB),
     Campus de la Universitad de las Islas  Baleares,
E-07122 Palma de Mallorca, Spain.
} 
\date{\today}

\begin{abstract}

We introduce a simple model of active transport  for an ensemble  of particles 
driven by an external shear flow. 
Active refers to the fact that the flow of the particles is modified by 
the distribution of particles itself.
The model consists in that the  effective velocity of every particle is
given by the average of the external flow velocities  felt by the particles
located at a distance less than a typical radius, $R$. Numerical analysis  reveals 
the existence of a transition to clustering  depending on the parameters of the
external flow and on $R$. A continuum description in terms of the 
number density of particles is derived, 
and a linear stability analysis of the density equation is performed 
in order to characterize the transitions observed in the model of interacting particles.

\end{abstract}
\maketitle



\section{Introduction}

Two different types of transport problems can be roughly distinguished: 
passive and active. The case of passive transport occurs when the
transported quantity does not affect the advecting flow, 
as exemplified by a dye immersed in a fluid, or of any 
reacting substance like a chemical pollutant having no feedback on the carrying atmospheric
or oceanic flow~\cite{general}. 
Conversely, in the active transport problem, the subject of this paper,
 the flow itself is modified by the advected
substance. Sometimes this is also refered as self-consistent transport since the velocity field
is in general determined by the substance via a dynamical constraint~\cite{diegochaos}.
 The temperature field,
an ensemble of charged particles moving in a self-generated electric field, the vorticity of a 
fluid flow, and gravitationally interacting particles, are a few examples reflecting 
the ubiquity and relevance of active transport processes in  Nature.

Recently, much progress has been achieved in both self-consistent and passive processes through
its reformulation within the Lagrangian description~\cite{falkovich, cencio}, which studies transport in terms
of individual particle trajectories instead of scalar fields. 
Thus, the Lagrangian description of a non-reacting passive scalar in an external
velocity field, $\vp (\bx,t)$, is given by
\BE
\frac{d\bx}{dt}=\vp (\bx,t)+\sqrt{2D_0}\etab (t),
\label{passive}
\EE
where $D_0$ is the diffusion coefficient of the passive scalar, and
$\etab$ is a normalized Gaussian white noise with zero mean and
delta correlated in time. In eq.~(\ref{passive}) the passive character is shown in the
fact that there is no coupling between the equations of motion. 
On the contrary, in active transport the interactions among particles 
alters the trajectory of any of them, so that for an ensemble of $N$ particles 
immersed in a fluid flow
one can write in general~\cite{boffetta}
\BE
\frac{d {\bf x}_{i}(t)}{d t} = {\vp}({\bx}_{1}(t), ..., {\bf x}_{N}(t)),
\label{active}
\EE
$i=1,...,N$.
This  $N$-body problem  is often treated in a mean-field approximation 
where every particle is considered independently of the rest but in an average potential 
determined self-consistently from the motion of all the particles~\cite{diegochaos, diego2}. 
In this way, the influence of any particle on the system is just through its contribution
to the potential. 

In this work, we introduce a distinct type of self-consistent transport model. 
At difference of the mean-field approach, our model
assumes a finite range of interaction, $R$,
for any particle, so that particles only interact with others surrounding them. 
Eq.~(\ref{active}) takes the form
\BE
\frac{d {\bf x}_{i}(t)}{d t} = {\vp}({\bx}_{i}(t), {\bx}_{i+1}(t) ..., {\bf x}_{i+N_R(i)}(t)),
\label{activemodelo}
\EE
where $N_R (i)$ is the number of particles at distance less than $R$ of particle $i$,
and with $i+1,..., i+N_R(i)$ we label these particles.
Most importantly, the external flow is {\it given} and the invidual particles modify their
response to the flow according to the local density around them. Like the mean-field one,
our model is an intermediate case between the passive transport case and the many-body
self-consistent models with real interactions decaying with distance. The aim
of the present paper is to show that even very simple active systems can show a 
very rich behaviour, and, in particular,
the formation of clusters of particles may appear. Also, due its simplicity one can present
a detailed analytical study of the model, and show that clustering emerges as a deterministic
instability of the density equation of the system.

The paper is organized as follows. In the next section we introduce the model and present numerical
results showing the clustering. Then, in Sec. III we derive the density equation
for the dynamics of the particles, and  perform a linear stability analysis of this continuum
description. Then, we finish in Sec. IV with the summary of the work.

\section{Self-Consistent model of particles driven by an external shear flow. Numerical results}

Let us consider $N$ particles in a two-dimensional system of size $L \times L$,
and  
the presence of a stationary incompressible two-dimensional  shear flow 
$\vp (x,y)=(0, v(x))$. 
In the  model  the effective velocity of  particle $i$ at time $t$,
$\vp^{eff}_i (t)$, which is in the position
$\bx_i (t)$,  is the 
average velocity  of the external velocities felt by the particles in its $R$-neighbourhood.
Mathematically
\begin{eqnarray}
\label{modelo1}
\vp^{eff}_i (t)&=&\frac{1}{N_R (i)}\sum_j \vp(\bx_j(t), t), \\
\frac{d\bx_i(t)}{dt}&=&\vp^{eff}_i (t),
\label{modelo2}
\end{eqnarray}
where, as indicated above,  $N_R (i)$ denotes the number of particles
 at a distance less than $R$ of particle $i$,
and the sum is restricted to the particles $j$ such that $|\bx_i (t)-\bx_j(t)| \le R$. 
Periodic boundary conditions are considered, and finite-size effects of the particles, like
inertia and collisions, 
are neglected.
Note that the self-consistent character of the model comes from the fact that at every
time the velocity of any particle is determined by the (local) distribution of particles
itself.
A noise term similar to eq.~(\ref{passive})
could be added to the r.h.s of eq.~(\ref{modelo1}), but this is not considered in this work and
we just suppose that advection induced by  the external flow dominates on the random motion
of the individual particles.

Two limits are clearly identified,  $R \to 0$   is the tracer limit, i.e., every particle is
simply driven by the flow. In the opposite $R \to L$ all the particles move with the same velocity,
 which is just an average of the external velocity field over all the  particles in the system. 
Physically, the model mimicks particles transported by a flow and 
 with some kind of effective non-local interaction 
that force them to move locally with the same velocity. In the context of living organisms, 
traffic or behaviour of human societies many different 
 models have been proposed where the density
of particles modify their velocity~\cite{flierl}:
 repulsion, attraction, distribution of resources, cooperation,
are some of the types  of interactions among the individuals that are usually studied. 
These interactions
are mediated (in a biological framework) through vision, hearing, smelling 
or other kinds of sensing, which is reflected, as in our model, by the appearence of a typical 
interaction radius,
$R$. However, a crucial difference of these biologically oriented models with
 (\ref{modelo1}-\ref{modelo2})
is that in those the particles are  self-propelled, i.e., they have their own velocity. In our model,
the velocity is externally given, and it is our aim to study the properties of the system of 
particles depending on the characteristics of the external flow.  Regarding
a biological motivation, our model is adequate for acuatic organisms
moving by the water flow that 
modify their velocity as a response  to other individuals living within a certain distance.

Concerning the clustering properties of the model, which is the main focus of this work,
 it is clear that 
shear enhances encounters among particles, and this   favours that particles
group together due to the averaging of velocities. On the contrary, local strength
of the external velocity field tends to  disperse the particles, breaking clusters. 
Combining these two effects a typical length scale is introduced:
\BE
\lambda^{-2} =\frac{<(dv(x)/dx)^2>}{<v(x)^2>}.
\label{escalataylor}
\EE
Here $<.>=1/L\int_0^L dx .$, and $\lambda$ is related to the Taylor microscale of turbulence, 
though here the meaning is somewhat different since it refers to the length scale at which
shear is comparable to the amplitud of the velocity.
 Therefore,  one expects the formation
of clusters when $\lambda$ is  smaller than the  typical interaction diameter, $2R$.
In other words, the typical length scale emerging from the comparison of shear and velocity must
be smaller than the scale at which we average the velocity of any particle.
 On the other
side, it is clear that when $R \approx L$ most of the particles of the system move with the
same velocity (all the particles enter in the average sum of (\ref{modelo1})), avoiding the
aggregation of the particles.
Thus our hypothesis for clustering requires that:
\BE
\lambda/2 \le R < L
\label{condition}
\EE

To be specific, in the following 
the external shear flow is 
 given by 
$v(x)=U_0 +V_0 \sin{(\omega x/L)}$, with $L$ the system size (which 
we take $L=1$ so that all length-scales are measured in units of $L$), $U_0$,
$V_0$ positive constants, and $\omega= 2\pi n$,  $n=0, 1, 2, ...$.  
For this flow it is not difficult to calculate
 $\lambda=\sqrt{1+2U_0^2/V_0^2}/\omega$
so that  on
taking $\sqrt{1+2 U_0^2/V_0^2}=2\pi$,
eq.~(\ref{condition}) becomes  $1/n \le 2R <1$. 

\begin{figure}
\begin{center}
\epsfig{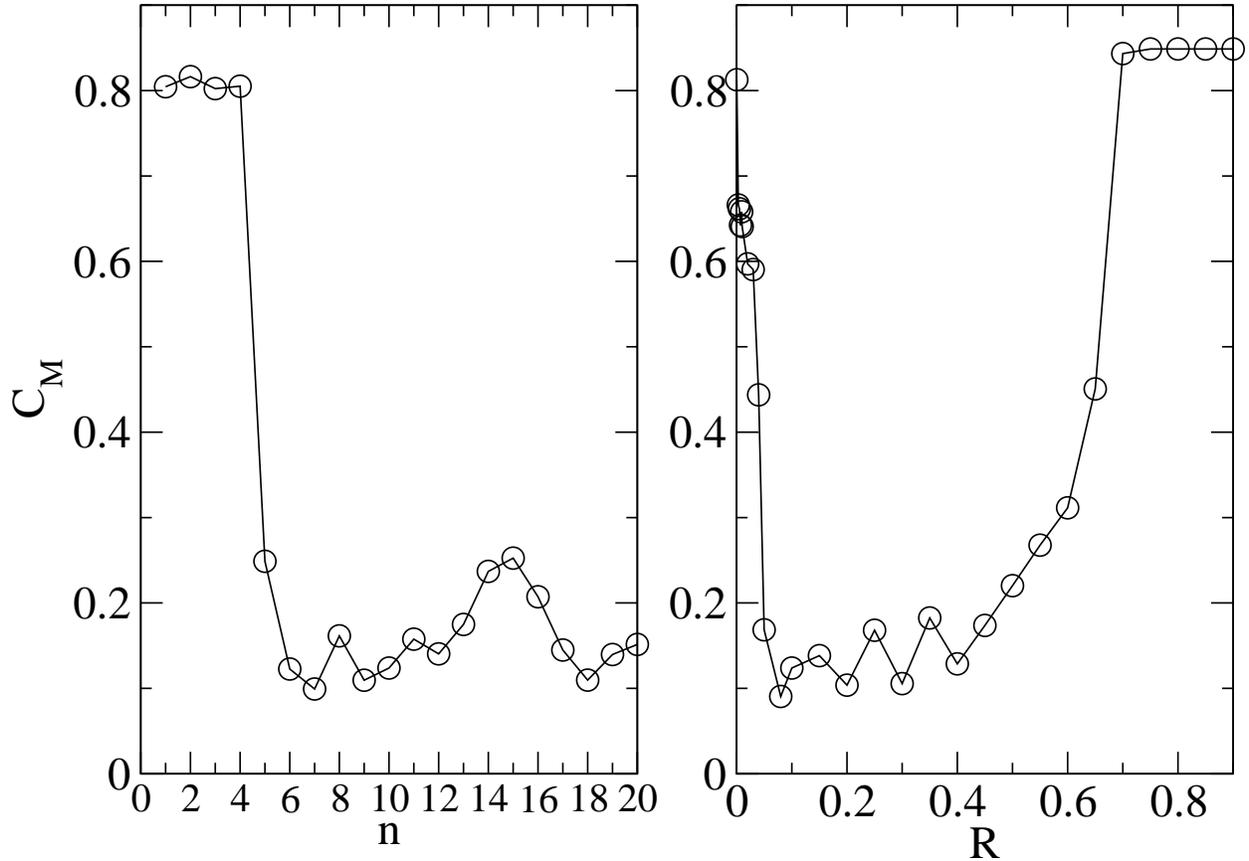}
\end{center}
\caption{Left: $C_M$ vs $n$ with $R=0.1$. Right: $C_M$ vs $R$ for $n=10$.
In both plots, $U_0=10$,  $\sqrt{1+2\frac{U_0^2}{V_0^2}}=2\pi$, and the
time average is performed over the last $2000$ steps in a numerical simulation running
for $5000$ steps with $dt=0.01$.
}
\label{fig:clustering1}
\end{figure}

\begin{figure}
\begin{center}
\epsfig{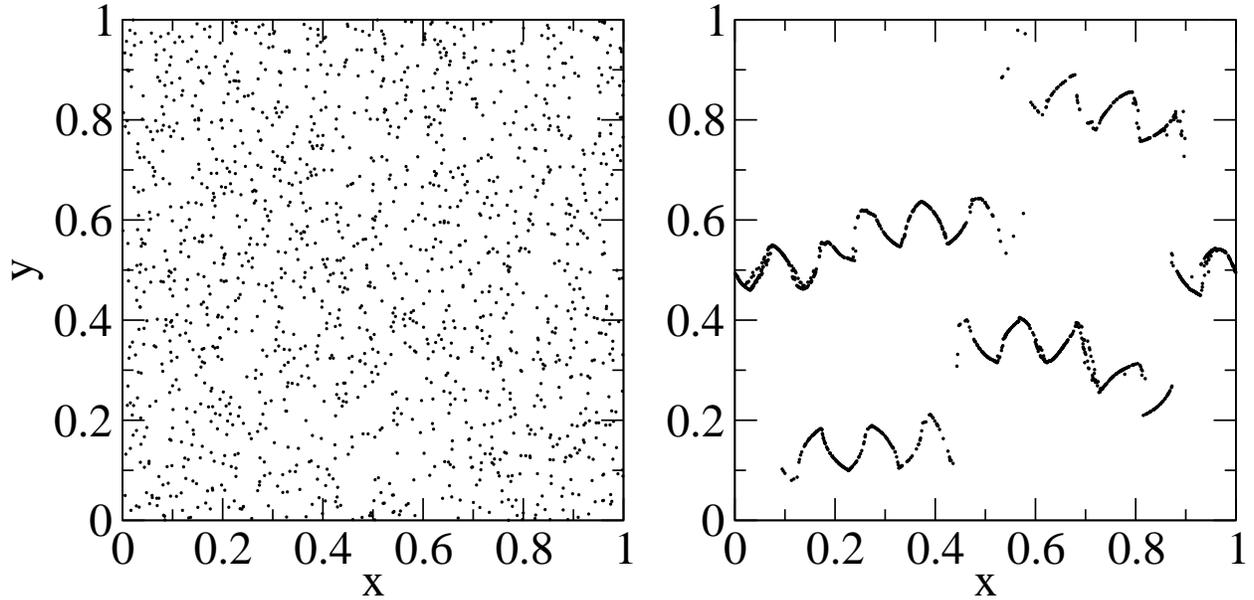}
\end{center}
\caption{Spatial distribution (statistically stationary) of particles at time $t=0$ (left),
and $t=16$ (right panel).
 Here $R=0.1$, $n=10$,
 $U_0=1$ ($ \sqrt{1+2\frac{u_0^2}{V_0^2}}=2\pi$, and the initial number of particles $N_0=1500$.
}
\label{fig:pattern}
\end{figure}

For a spatial distribution of particles 
the quantitative characterization of clustering \cite{puglio}
is performed by means of an  entropy-like measure
\BE
S_M=-\sum_{i=1}^M \frac{m_i}{N}\ln \frac{m_i}{N},
\label{entropy}
\EE
where $M$ is the number of boxes in which we divide the system, and $m_i$ is the number of particles
in box $i$. One has that 
$0\le S_M \le \ln M$, such that $S_M=0$ is obtained when all the particles are
in just one of the boxes, and the $\ln M$ value is reached
when $m_i= N/M$ for all $i$ (Poisson 
distribution of particles), i.e., $S_M$ decreases when the clustering increases. We define
the clustering coefficient as $C_M=\exp(<H_M>_t)/ M$, where $<.>_t$ denotes a temporal average at long
times, so that when there is no clustering $C_M \approx 1$. 
In the left panel of  fig. (\ref{fig:clustering1})   we fix $R=0.1$ (much smaller than 
the system size $L=1$) and plot $C_M$ vs $n$ observing
that the transition to clustering is obtained for $n \approx 5$, fitting perfectly (\ref{condition}).
In the right panel we take $n=10$ and plot $C_M$ vs $R$ observing the two transitions indicated
in (\ref{condition}). 
In figure~\ref{fig:pattern} we plot 
the spatial distribution of particles (in the left panel we plot the initial distribution)
in the regime of clustering
at time  $t=16$ (right) for $R=0.1$ and $n=10$. Here one sees that the particles
tend to aggrupate  following the sinusoidal
flow.

Similar results are obtained for other shear flows. E.g., for the linear shear given by 
$v(x)=\Gamma x$ if $x\in [0, 1/2]$ and $v(x)=\Gamma (1-x)$ for $x\in [1/2, 1]$, 
$\lambda$ 
  is $1/\sqrt{12}$ so that altering the features of the external flow 
(the values of $\Gamma$)  the aggregation properties of the system for fixed $R$
are not changed. 
However, transitions between non-clustering and clustering distributions are observed by
varying $R$.

In the next section we explain analytically the transition to clustering observed in the numerics.
This is done by deriving the density evolution equation for the system of particles.

\section{Continuum description in terms of the density of particles. Linear
stability analysis}

A continuum theory  can give further insight on the model. 
The process to obtain it is standard \cite{Dean}, and we just present here a sketch:
define the particle
density as $\rho (\bx, t)=\sum_{i=1}^N \rho_i (\bx, t)=\sum_{i=1}^N \delta (\bx_i (t)-\bx)$,
then use  an arbitrary function $f(\bx)$ defined on the coordinate space, and take the
time derivative on both sides of the obvious 
relation $f(\bx_i (t))=\int d\bx \ \rho_i (\bx, t) f(\bx )$.  Finally, using 
$\int_{|\bx - \bx_i (t)|\le R} d\bx \ \rho (\bx, t)= N_R (i)$ one arrives to
\BE
\partial_t \rho (\bx, t)+\nabla_{\bx} \cdot 
\left[\frac{\rho (\bx, t) \int_{|\br -\bx|\le R} d\br\ \vp (\br, t)\ \rho(\br, t)}
{\int_{|\br -\bx|\le R} d\br\ \rho(\br, t)}\right]=0.
\label{continua}
\EE
Note that we have maintained the time dependence of the velocity field to reflect the
generality of the approach.
Eq.~(\ref{continua}) can be simply read as that the density of particles is driven by
the effective velocity $\vp^{eff} (\bx, t)= \int_{|\br -\bx|\le R}
 d\br\ \vp (\br, t)\ \rho(\br, t)/\int_{|\br -\bx|\le R} d\br\ \rho(\br, t)$, whose
dependence on the density reveals the self-consistent character of the model. Note also
the two trivial limits: a) $R \to 0$ or passive limit, $\vp^{eff} (\bx, t) \to \vp (\bx, t)$, and
b) $R \to 1$ ($L=1$) for which $\vp^{eff} (\bx, t) \to 1/N\int d\br \rho(\br, t) \vp (\br, t) $,
 i.e., the average velocity 
of the system of particles, which is the same for all of them (and constant for a time-independent
velocity field).

Next we make a linear stability analysis of  the stationary homogenous 
solution, $\rho_0$, of eq.~(\ref{continua}). We first write 
$\rho (\bx, t)=\rho_0 + \epsilon \psi (\bx, t)$ where $\epsilon$ is a small parameter, 
and $\psi (\bx, t)$ the space-time dependent perturbation, and substitute it in 
eq.~(\ref{continua}). To first order in $\epsilon$, using incompressibility
of the flow and denoting $\int_B . =\int_{|\br -\bx|\le R} .$  we obtain
\begin{eqnarray}
&\partial_t \psi+\frac{1}{\pi R^2}\bg (\bx)\cdot\nabla_{\bx}\psi
+\frac{1}{\pi R^2}\nabla_{\bx} \cdot \int_B d\br \ \vp (\br, t) \psi (\br, t)& \nonumber \\
&-\frac{1}{(\pi R^2)^2}\bg (\bx)\cdot \int_B d\br \  \psi (\br, t) =0,&
\label{monstruo}
\end{eqnarray}
with $\bg (\bx)=\int_B d\br \ \vp (\br, t)$. Though linear, the above expression
is still rather complicated
since it is non-local in space. For the sinusoidal shear flow
(taking for simplicity and without lost of generality $U_0=0$),
we have that $\bg (\bx)=\hat \by 2\pi  R/\omega  \sin (\omega x) J_1 (\omega R)$
with $\hat \by$ a unitary vector in the $y$ direction, and $J_1$ the first order Bessel function,
so that
eq.~(\ref{monstruo})  becomes
\begin{eqnarray}
&\partial_t \psi+\frac{2 V_0 J_1(\omega R)}{\omega R}\sin (\omega x)
\partial_y \psi& \nonumber \\
&+\frac{V_0}{\pi R^2}\partial_y
[\int_B d\br \ \sin (\omega r_x)\psi (r_x, r_y, t)]& \nonumber \\
&-\frac{2 V_0 J_1 (\omega R)}{\pi \omega R^3}\sin (\omega x)\partial_y[
\int_B d\br \  \psi (\br, t)] =0,&
\label{monstruosin}
\end{eqnarray}
where $\br=(r_x, r_y)$.

We are mainly interested in the clustering transition driven by the  relative values of 
$\lambda$ and $R$, so that we next consider the limit $R << 1$. It is  very important
to note that, to be the  expansion in $R$ consistent, all length-scales of the system must
also be very small compared with the system size. Specifically, $\lambda=1/\omega <<1$, so that
in particular, one cannot expand the $\sin (\omega r_x)$ in the integrand in eq.~(\ref{monstruosin}).
Let us detail the calculations. The two integrals appearing  in eq.~(\ref{monstruosin}) 
have the approximations (for simplicity of notation we skip the time dependence):
\begin{eqnarray}
&I_a=\int_B d\br \psi (r_x, r_y)= \int_{|\br'|\le R} d\br' 
 \psi (r_x'+x, r_y'+y) &  \nonumber \\
&\approx \int_{|\br'|\le R} d\br'[\psi(x, y) +r_x' \partial_y \psi (x, y)
+r_y' \partial_y \psi (x, y)]&\nonumber \\
&=\pi R^2  \psi (x, y) +\Theta (R^4);& 
\label{exp1} \\
&I_b=\int_B d\br \sin (\omega r_x) \psi (r_x, r_y)&\nonumber \\
&=\int_{|\br'|\le R} d\br'\sin (\omega r_x'+\omega x) \psi (r_x'+x, r_y'+y)& \nonumber \\
&\approx \int_{|\br'|\le R} \sin (\omega r_x'+\omega x) 
[\psi(x, y) +r_x' \partial_y \psi 
+r_y' \partial_y \psi ]& \nonumber \\
&= \frac{2\pi J_1 (\omega R) R}{\omega}\sin (\omega x) \psi + 
\frac{4  \pi R^2 J_2 (\omega R)}{\omega}\cos  (\omega x) \partial_x \psi +\Theta(R^4). & \nonumber \\
\label{exp2}
\end{eqnarray}
Here $\Theta (R^4)$ indicates terms of order $R^4$ and superior.
After substituting expressions (\ref{exp1})-(\ref{exp2}) in eq.~(\ref{monstruosin}) the evolution of 
the perturbation in the small $R$ limit (or better, when the typical length scales of the problem
are small) is finally given by
\begin{equation}
\partial_t \psi+\frac{2 V_0 J_1(\omega R)}{\omega R}\sin (\omega x)
\partial_y \psi
+\frac{4 V_0 J_2 (\omega R)}{\omega}\cos (\omega x)\partial_{xy}^2 \psi  =0,
\label{definitivaper}
\end{equation}
where we have neglected terms of order $R$.


Two
fundamental features further
simplifies the analysis: a) the coefficients are periodic in the spatial coordinates
so that Floquet theory can be applied, and b) the coefficients are independent of the
$y$ coordinates so that plane waves are solutions on the $y$ direction. Therefore we make the 
ansatz:
\BE
\psi (x, y, t) = e^{\Lambda t +i \hat \omega y +i K x}\sum_{m=-\infty}^{\infty} \phi_m e^{i\omega x m},
\label{ansatz}
\EE
where, because of periodic boundary conditions, $\hat \omega =2 \pi p$ ($p=1, 2, ...$), 
$K=2 \pi p$ ($p=1, 2, ...$), and 
$\phi_m$ are complex coefficients. 
$K$ is  restricted to the first Brillouin zone determined by $-\omega/2 \le K \le \omega/2$,
and $\hat \omega$ is not bounded.

If any of the eigenvalues $\Lambda$ is positive then
the perturbation grows (the homogenous solution is unstable) and clustering  emerges in the system.
Thus we look for the conditions to have $\Lambda >0$. Using the exponential formula for 
the sine and cosine functions, and substituting expression (\ref{ansatz}) in
eq.~(\ref{definitivaper}) we obtain, after grouping the terms with the same exponential argument,
\begin{equation}
\Lambda_m \phi_m +
\phi_{m-1}[\alpha_1 -\beta m]
+\phi_{m+1} [\alpha_2 -\beta m ]=0,
\label{fourier}
\end{equation}
with $\alpha_1=a_1 \hat \omega /2 -a_2 \hat \omega K/2 +a_2 \omega  \hat \omega/2 $,
$\alpha_2= -a_1 \hat \omega /2 -a_2 \hat \omega K/2-a_2 \omega  \hat \omega/2$,
and
$\beta=a_2 \omega \hat \omega /2 $, with the notations $a_1=2 V_0 J_1(\omega R)/(\omega R)$
and $a_2=4 V_0 J_2 (\omega R)/\omega$.

 For a simple theoretical analysis we just consider
the three Fourier modes $m=0, \pm 1$ and neglect the rest. Diagonalizing the
corresponding 
$3 \times 3$ matrix of coefficients of the 
system in eq.~(\ref{fourier}) we obtain three eigevanlues, one  zero and the other two 
given by 
\begin{equation}
\Lambda_{\pm} (K) = \pm
\frac{V_0\hat \omega}{\omega R} 
\sqrt{8 J_2^2 K^2 R^2 -4 \omega R J_2 J_1 -2 J_1^2},
\label{eigenvalue}
\end{equation}
where the Bessel functions, $J_1$ and $J_2$, are evaluated at $\omega R$.
The expression for $\Lambda_+$ is cuadratic in $K$ with a positive coefficient for
the term in $K^2$,
so that taking into account that $-\omega/2\le K \le \omega/2$, the
inestability is obtained when  $\Lambda_+ (K=\omega /2)$ is positive, i.e.,
\begin{equation}
J_2(\omega R)^2 \omega^2 R^2 - 2 \omega R J_2(\omega R) J_1 (\omega R)- J_1 (\omega R)^2 \ge 0.
\label{condition3modos}
\end{equation}
Numerically one solves the above inequality and obtains that the condition for instability
is  $\omega R \ge 2.5$, 
which, despite the many approximations made to derive it, fits well with the numerical result
$1/\omega \le 2 R$ (eq.~(\ref{condition}) for $U_0=0$). We have checked that the above
result is  improved by including more modes in the Floquet analysis. In particular,
considering $m=0, \pm 1, \pm 2$, the final condition for the maximum exponent  to be positive
is $\omega  R \ge 1.32 $. We believe that in the limit $m \to \infty$ the numerical result is approached.
Therefore this analysis confirms that the derived continuum description eq.~(\ref{continua}) properly
describes the discrete interacting particles model, and that the clustering emerges as a
deterministic instability of the density equation.

\section{Summary}

In this work we have proposed a very simple model for  an ensemble of particles self-consistently
driven by an external shear flow. Despite its simplicity the model shows a very interesting
behavior where a transition to grouping of particles are observed. An hypothesis for the
appearence of the clustering has been presented. It esentially says that the clustering 
appears when the length scale that comes from the comparison of the shear flow  and
the velocity field amplitudes is smaller than the typical interaction radius of the particles. 
This hypothesis has been numerically checked and also a continuum description has been derived that
confirms it.

A more realistic interaction of the particles, for example decaying with the distance within
$R$, is planned
to be studied in the future. Also, it will be interesting 
a detailed study of the role of a noise term in the dynamics of the particles
(which in the continuum description is a difussion term), and the analysis 
when a chaotic flow is considered.

\section{Acknowledgments}

I have benefited from many useful conversations  with Emilio Hern\'andez-Garc\'\i a. 
I also acknowledge discussions with Dami\`a Gomila and Pere Colet.
Work  supported 
from MCyT of Spain under projects
REN2001-0802-C02-01/MAR (IMAGEN) and BFM2000-1108 (CONOCE), and from a  
 {\it Ram\'on y Cajal}
fellowship of the Spanish MEC.

\end{document}